\documentclass{ieeeaccess}
\usepackage{cite}
\usepackage{amsmath,amssymb,amsfonts}
\usepackage{algorithmic}
\usepackage{graphicx}
\usepackage{comment}
\usepackage{textcomp}
\def\BibTeX{{\rm B\kern-.05em{\sc i\kern-.025em b}\kern-.08em
    T\kern-.1667em\lower.7ex\hbox{E}\kern-.125emX}}
\begin{document}
\history{Date of publication xxxx 00, 0000, date of current version xxxx 00, 0000.}
\doi{10.1109/ACCESS.2017.DOI}

\title{Influence of anthropomorphic agent on human empathy through games}
\author{\uppercase{Takahiro Tsumura}\authorrefmark{1}\authorrefmark{2}, 
\uppercase{Seiji Yamada\authorrefmark{2}}\authorrefmark{1}}
\address[1]{The Graduate University for Advanced Studies, SOKENDAI, Tokyo, Japan}
\address[2]{National Institute of Informatics, Tokyo, Japan}
\tfootnote{This work was partially supported by JST, CREST (JPMJCR21D4), Japan.
This work was also supported by JST, the establishment of university fellowships towards the creation of science technology innovation, Grant Number JPMJFS2136.}

\markboth
{Takahiro Tsumura \headeretal: Anthropomorphic agent influences human empathy through task intervention}
{Takahiro Tsumura \headeretal: Anthropomorphic agent influences human empathy through task intervention}

\corresp{Corresponding author: Takahiro Tsumura (e-mail: takahiro-gs@nii.ac.jp).}

\begin{abstract}
The social acceptance of AI agents, including intelligent virtual agents and physical robots, is becoming more important for the integration of AI into human society. 
Although the agents used in human society share various tasks with humans, their cooperation may frequently reduce the task performance. One way to improve the relationship between humans and AI agents is to have humans empathize with the agents. 
By empathizing, humans feel positively and kindly toward agents, which makes it easier to accept them. 
In this study, we focus on tasks in which humans and agents have various interactions together, and we investigate the properties of agents that significantly influence human empathy toward the agents. 
To investigate the effects of task content, difficulty, task completion, and an agent's expression on human empathy, two experiments were conducted. 
The results of the two experiments showed that human empathy toward the agent was difficult to maintain with only task factors, and that the agent's expression was able to maintain human empathy. 
In addition, a higher task difficulty reduced the decrease in human empathy, regardless of task content. 
These results demonstrate that an AI agent's properties play an important role in helping humans accept them.
\color{black}
\end{abstract}


\titlepgskip=-15pt

\maketitle

\section{Introduction}
\label{sec:introduction}
\PARstart{A}{nthropomorphic} agents are increasingly being utilized in human society. These can be considered a kind of tool used by humans, and historically, tools have sometimes been treated as if they are actually sentient. For example, humans have treated artificial objects as though they were human in media equations \cite{Reeves96}. Not all of the artifacts in use today are perceived in this way, however. The main problems currently facing AI implementation are related to reliability and ethical usage. One recent study on trust, AI ethics, and the anthropomorphizing of AI argued that even complex machines should not be viewed as trustworthy \cite{Ryan20}; instead, people should ensure that the organizations using AI and the individuals within those organizations are trustworthy. AI ethics was also discussed in depth from an applied-ethics perspective in a study by Hallamaa and Kalliokoski \cite{Hallamaa22}.
\\ \indent
Empathy for artifacts can occur when humans treat them as if they were alive. Artificial objects that we empathize with include cleaning robots, pet-type robots, and anthropomorphic agents that provide services such as online shopping and help desk assistance. The appearance of these agents varies depending on the application and environment. Although they are already used in society and coexist with humans, some people simply cannot accept this type of agent \cite{Nomura06,Nomura08,Nomura16}. As anthropomorphic agents are predicted to be utilized in human society even more in the future, they need to have functions that are acceptable to humans.
\\ \indent
In this study, we focus on how to help humans improve their relationships with agents. To this end, we developed an empathy agent to induce human empathy, where ``empathy" is an attribute that the agent should have in order to be accepted by human society. Our concept is that, by empathizing, people will act more positively toward agents and be more likely to accept them. Various studies have investigated what induces empathy, including those that focus on potential factors such as linguistic information \cite{Konrath15,Shaffer19,Tahara19}, nonverbal information \cite{Tisseron15,Yoshioka15,Okanda19}, situations \cite{O'Connell13,Zhi18,Richards18}, and relationships \cite{Stephan15,Hosseinpanah18,Giannopulu20}.
\\ \indent
In our approach, we assume that the empathy agent itself is what influences human empathy. Our main focus is the factors that make agents acceptable to humans, so we investigate only human empathy toward agents, and consider the human capacity for empathy to be outside the current scope. 
\\ \indent
To this end, we perform two experiments to examine the impact of the task on human empathy for the agent and to investigate whether the agent changes human behavior.
\color{black}
This is an important study that investigates the influence of factors to improve the human-agent relationship. 
In this paper, we propose our empathy agents for promoting human empathy. Then, we discuss our experiments and method. Finally, we discuss the results and describe our future work.

\section{Related work}
\subsection{Definition of empathy}
We consider empathy to be a vital element in the human acceptance of agents as members of society. This mirrors the importance of humans empathizing with each other in order to get along \cite{Gaesser13,Klimecki16}. 
\\ \indent
Empathy and the influence it has on others has been studied extensively in the field of psychology. Omdahl \cite{Omdahl95} classified empathy into three types: (1) affective empathy, which is an emotional response to the emotional state of others, (2) cognitive empathy, which is a cognitive understanding of the emotional states of others, and (3) empathy that includes both of the previous types. Preston and de Waal \cite{Preston02} suggested that at the core of the empathic response is a mechanism that enables the observer to access the subjective emotional state of the target. They came up a perception-action model (PAM) to unify the different perspectives in empathy and classified empathy into three types: (a) sharing or being influenced by the emotional state of others, (b) assessing the reasons for the emotional state, and (c) having the ability to identify and incorporate other perspectives. Olderbak et al. \cite{Olderbak14} examined theoretical and empirical support for the emotional specificity of empathy and proposed an emotion-specific empathy questionnaire that assesses affective and cognitive empathy for six basic emotions.
\\ \indent
Our study focuses on the effect of empathy in promoting society's acceptance of anthropomorphic agents, but in psychology, empathy has been explored from a variety of other aspects, including its negative effects. Bloom tried to introduce a neutral aspect of empathy that encompasses not only positive influences but also negative ones \cite{Bloom16}. He claimed that when empathy acts as a moral trigger, it can lead to humans making irrational decisions and to relationships that erupt into violence and anger. He also suggested that we can overcome this negative aspect by combining conscious and deliberative reasoning with an altruistic approach.
\\ \indent
Various questionnaires have been utilized to measure empathy. The Ten Item Personality Inventory (TIPI) is one of the more popular ones, but since it investigates human personality as a whole \cite{Gosling03}, it is unsuitable for our purposes because we presume empathy is biased by human personality. The Interpersonal Reactivity Index (IRI), also commonly used in the field of psychology, is designed to investigate the specific characteristics of empathy \cite{Davis80}. Baron-Cohen and Wheelwright \cite{Baron-Cohen04} proposed a self-report questionnaire called the Empathy Quotient (EQ) for use with adults of normal intelligence. Lawrence et al. \cite{Lawrence04} investigated the reliability and validity of the EQ and found that there is a moderate association between the EQ subscale and the IRI subscale. We ultimately opted to use only IRI as the questionnaire for our experiment.

\subsection{Empathy in engineering}
Empathy has also been studied in the field of engineering, particularly in the context of virtual reality. A study by van Loon et al. \cite{vanLoon18} investigated whether the effects of VR perspective-taking could be driven by increased empathy and extended to real-stakes behavioral games. They succeeded in increasing the tendency of participants to take the other person's point of view, but only if it was that of the same person participants assumed in the virtual reality simulation.
\\ \indent
Herrera et al. \cite{Herrera18} compared the short- and long-term effects of traditional and VR viewpoint acquisition tasks. They also conducted experiments investigating the role of technological immersion with respect to different types of intermediaries. Crone and Kallen \cite{Crone22} utilized online platforms and immersive virtual reality to examine the role of virtual perspective-taking on binary gender. They found that virtual reality-based perspective-taking may have a greater impact on acute behavioral modulation of gender bias compared to online because it immerses participants in the experience of temporarily becoming another.
\\ \indent
Empathy is also attracting attention in the realm of product design. Bennett and Rosner \cite{Bennett19} investigated a human-centered design process (promise of empathy) in which designers try to understand the target user with the aim of informing technology development. Rahmanti et al. \cite{Rahmanti22} designed a chatbot with artificial empathic motivational support for dieting called ``SlimMe'' and investigated how people responded to the diet bot. They proposed a text-based emotional analysis that simulates artificial empathic responses to enable the bot to recognize users' emotions.
\\ \indent
In the fields of human-computer interaction (HCI), empathy between humans and agents or robots is studied. 
Wright and McCarthy \cite{Wright08} discussed the use of empathy, citing studies that have used empathy in HCI. 
Pratte et al. \cite{Pratte21} analyzed 26 publications on empathy tools and developed a framework for empathy tool designers.

\subsection{Empathy in human-robot interaction}
Studies on human empathy in the human-robot interaction (HRI) field have explored the ways in which humans empathize with artificial objects. 
\\ \indent
Beck et al. \cite{Beck10} investigated the effect of changing a robot's head position on the interpretation of emotional key poses, valence, arousal, and stances. Their findings support the idea that body language is an appropriate medium for robots to express emotions. Hofree et al. \cite{Hofree14} showed that human participants spontaneously matched the facial expressions of androids in the same room. This mimicry made the participant feel uncomfortable toward the android, even though the participant was fully aware of the android's lack of ill intent. This result suggests that mimetic responses depend on the prominence of humanlike features emphasized by face-to-face interaction, thus highlighting the role of presence in human-robot interactions.
\\ \indent
On the basis of the concept of cognitive developmental robotics, Asada \cite{Asada15} proposed ``affective developmental robotics" as a way to produce more authentic artificial empathy. Artificial empathy here refers to AI systems (e.g., companion robots and virtual agents) that can detect emotions and respond empathetically. The design of artificial empathy is one of the most essential components of social robotics, and empathetic interaction with the public is necessary to introduce robots into society.
\\ \indent
Fraune \cite{Fraune20} examined how people behave morally and perceive players according to their group membership (in-group, out-group), agent type (human, robot), and robot anthropomorphism (anthropomorphic, mechanized). Their results showed that the pattern of reactions to humans was more favorable for anthropomorphic robots than for mechanistic robots. Tuyen et al. \cite{Tuyen21} designed emotional bodily expressions for a robot that adopts the user's emotional gestures. They proposed an action selection and transformation model that enables the robot to progressively learn from the user's gestures, identify the user's habitual behaviors, and transform the selected behaviors into robot actions. Their findings showed that the robot's emotional expressions reflecting the characteristics of the partner were widely accepted within the same cultural group and perceived in different ways among different cultural groups.

\subsection{Empathy in human-agent interaction}
As for empathy research in the field of human-agent interaction (HAI), Leite et al. \cite{Leite14} conducted a long-term study in an elementary school where they presented and evaluated an empathy model for social robots aimed at interactions with children that occur over a long period of time. They measured children's perceptions of social presence, engagement, and social support and found that the empathy model developed had a positive impact on the long-term interaction between the child and the robot. Deshmukh et al. \cite{Deshmukh18} analyzed the relationship between the way robots gesture and the way those gestures are perceived by human users. In particular, they investigated how changing the amplitude and speed of a gesture affects the Godspeed score given to that gesture. Their results suggested that forming gestures aimed at making explicit the internal state of the robot tended to change the perception of animacy, while forming gestures aimed at interaction effects tended to change the perception of anthropomorphism, desirability, and perceived safety.
\\ \indent
Chen and Wang \cite{Chen19} hypothesized that empathy and anti-empathy were closely related to a creature's inertial impression of coexistence and competition within a group and established a unified model of empathy and anti-empathy. They also presented the Adaptive Empathetic Learner (AEL), an agent training method that enables evaluation and learning procedures for emotional utilities in a multi-agent system. Perugia et al. \cite{Perugia20} investigated which personality and empathy traits were related to facial mimicry between humans and artificial agents. They focused on the humanness and embodiment of agents and the influence that these have on human facial mimicry. Their findings showed that mimicry was affected by the embodiment that an agent has, but not by its humanness. It was also correlated with both individual traits indicating sociability and empathy and traits favoring emotion recognition.
\\ \indent
Tsumura and Yamada \cite{Tsumura22} investigated empathy as one way to improve the relationship between humans and anthropomorphic agents. They focused on tasks between agents and humans and experimentally examined the hypothesis that task difficulty and task content promote human empathy. Their findings showed that there was no main effect for the task content factor but a significant main effect for the task difficulty factor.
\\ \indent
To clarify the empathy between agents/robots and humans, Paiva represented the empathy and behavior of empathetic agents (called empathy agents in HAI and HRI studies) in two different ways: targeting empathy and empathizing with observers \cite{Paiva04,Paiva11,Paiva17}. 
In our study, following Paiva's proposed definition of an empathic agent, we consider the agent as an object of empathy and examine how the empathy and empathic responses of the human participants are affected.
\color{black}

\subsection{Prior research on the task}
A study of cooperative and competitive tasks was conducted by Ruissen and de Bruijn \cite{Ruissen16}. 
In the study, cooperative and competitive tasks were tested using Tetris. 
The results confirmed that the cooperative task did not reduce self-integration, but the competitive task did. 
Another study of competitive tasks between humans and robots is that of Kshirsagar et al. \cite{Kshirsagar19}. 
They performed a human-robot competitive task using the same task and found that participants preferred a lower-performing robot to a higher-performing one.
 Boucher et al. \cite{Boucher12} performed a human-robot cooperation task. 
The robot recognized gaze guidance to the human faster than the robot gave voice instructions to the human.
\\ \indent
Some studies of task difficulty include the following. 
Fuentes-García et al. \cite{Fuentes19} used chess problem-solving tasks of different difficulty levels to investigate participants' heart rate variability in terms of difficulty, stress, complexity, and cognitive needs. 
Cho \cite{Cho21} considered that task difficulty and mental workload are necessary to improve the usability and frequency of use of interactive systems and proposed a new approach for automatically estimating task difficulty by focusing on human blinking.

\section{Study 1 method}
In this study, we focused on a task between agents and humans. 
We experimentally investigated hypotheses stating that task difficulty and task content facilitate human empathy. 
The experiment was a two-way analysis of variance (ANOVA) with four conditions: task difficulty (high, low) and task content (competitive, cooperative). 

\subsubsection{Experimental goals and design}
The purpose of this study is to investigate whether task difficulty and task content can elicit more human empathy in an interaction with an empathy agent.
We are the first study to relate task difficulty and task content to empathy and apply it to HAI. 
This research will facilitate the application of agents used in human society by influencing human empathy. 
In addition, if there is a change in human empathy due to the influence of a task, the importance of the task can be discussed among humans. 
For these purposes, we developed two hypotheses.
\begin{enumerate}
\item[\textbf{H1}:] When performing a competitive task with an empathy agent, the higher the task difficulty, the more human empathy is suppressed.
\item[\textbf{H2}:] When cooperating with an empathy agent, the higher the task difficulty, the more human empathy is promoted.
\end{enumerate}

The above hypotheses were determined by inferring from the results of the Ruissen and de Bruijn \cite{Ruissen16} and Fuentes-García et al. \cite{Fuentes19} studies.
The above hypotheses were reached because, in cooperative tasks, humans improve their performance and have favorable impressions of their cooperating partners, whereas in competitive tasks, they think less about their adversaries. 
\\ \indent
In addition, as empathy changes with task difficulty, the higher the difficulty, the greater the mental load, and the greater the impact on performance. 
Therefore, we hypothesized that, in competitive tasks, the higher the task difficulty, the more human empathy would be suppressed. 
\\ \indent
In comparison, in the cooperative task, task performance was improved by quickly reading the intentions of the cooperating partner, which may be related to perspective taking in the cognitive empathy category. 
Since the task was facilitated by putting oneself in the other person's shoes and reading the other person's actions, we hypothesized that human empathy is facilitated in cooperative tasks as the task difficulty increases.
\\ \indent
An experiment was conducted to investigate these hypotheses with a two-factor between-participants design with two factors: task difficulty and task content. 
The number of levels for each factor was two for difficulty (high, low) and two for content (competitive, cooperative). 
Participants took part in only one of four different content conditions. 
The dependent variable was the empathy held by the participants.
Figures 1 and figure 2 show this experiment.

\Figure[tbp][scale=0.3]{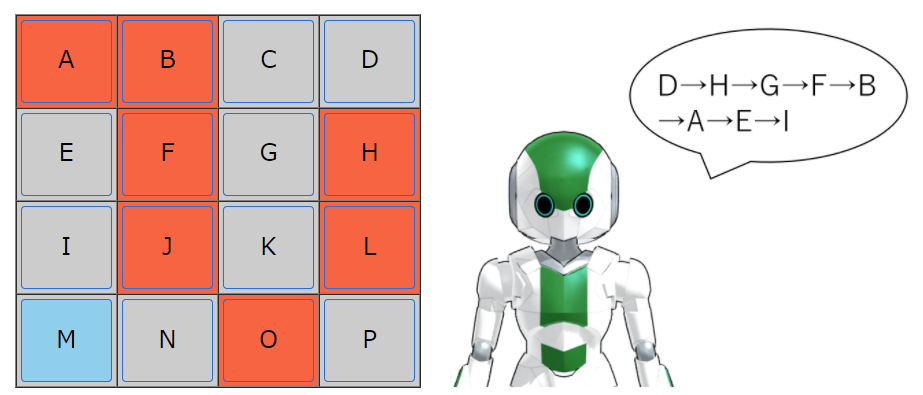}{Task scene with empathy agent during high difficulty\label{fig1}}

\begin{figure}[tbp]
		\begin{center}
		\includegraphics[scale=0.3]{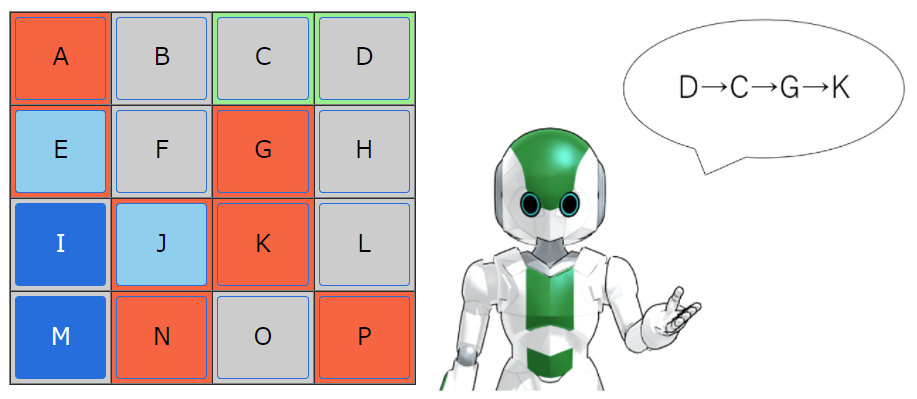}
		\caption{Task scene with empathy agent during low difficulty}
		\label{fig2}
	\end{center}
\end{figure}

\subsubsection{Experimental details}
The experiment was conducted in an online environment. 
The online environment used has already become a common method of experimentation \cite{Davis99,Crump13,Okamura20}. 
Figure 3 shows the flowchart of the experiment.

\begin{figure*}[tbp]
\centering
    \includegraphics[scale=0.35]{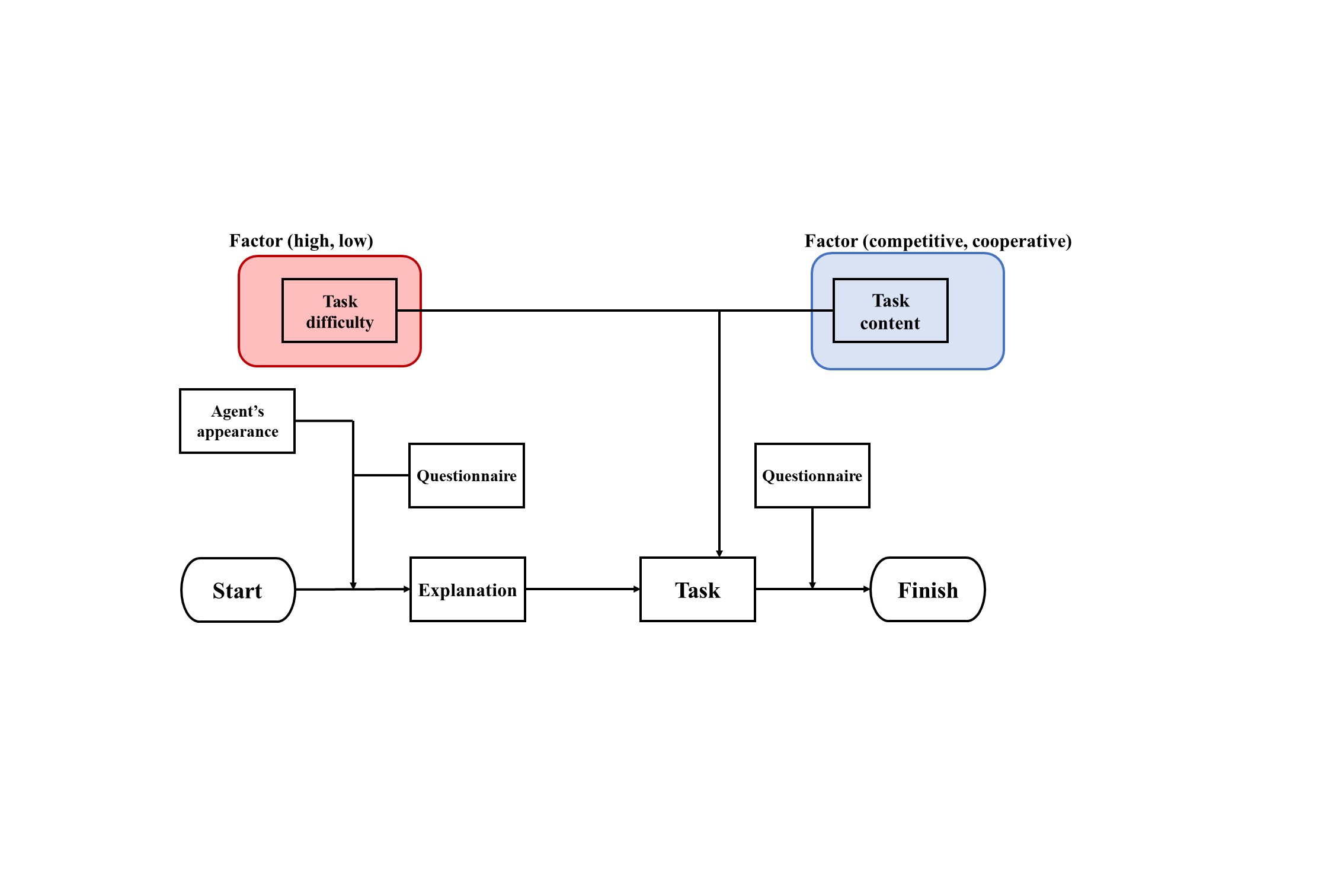}
    \vspace{1mm}
    \caption{Flowchart of the experiment.}
    \label{fig3}
\end{figure*}

As mentioned, the purpose of this study is to promote human empathy toward anthropomorphic agents. 
When performing a task with an anthropomorphic agent, the environment is assumed to be accessed via a PC rather than in reality, so we thought that the same effect could be achieved even with an online environment.
\\ \indent
Before performing the task, a questionnaire was administered to measure the empathy toward the empathy agent. 
At this point, participants were not allowed to make judgments about the task difficulty or task content. 
The reason for administering the questionnaire before the task was to check for differences in participants' empathy toward the agent.
In the flowchart of Figure 3, the participants were asked to answer questions before the explanation.
Thus, the pre-task question cannot be used as a factor because it is before the task as interaction between the agent and participants.
\color{black}
\\ \indent
Tasks were set up differently depending on the content of the competition and cooperation, but to avoid drastic changes in task content, the common denominator was to move between squares within a range of 16 (4 × 4 squares). 
Common to all tasks was that participants always moved from the bottom left square, and the agent always moved from the top right square. 
The mass movement were above, below, to the left, and to the right of the current point, and the same square could only be passed through once. 
The total number of moves that could be made alternated, but the total number for the agent was indicated at the start. 
\\ \indent
The reason why this information for the agent was given advance is that it increased the difficulty of the task and reduced the difference in difficulty between the two difficulty levels. 
If this information were not given, participants would need to think about how many moves the agent could make, which would be burdensome and would affect the judgment of the difficulty level. 
The task was made simple so that the comparison of difficulty levels would not be affected by the selections of the agent while trying to anticipate their movement total. 
Every task had a checkpoint, and the purpose was to pass through this point. 
The squares selected by the participants were colored blue, the squares that could be moved to next were colored light blue, the squares selected by the agent were colored light green, and the checkpoints were colored vermilion.
\\ \indent
The task was performed three times in total, with different checkpoint locations. 
The total number of moves varied depending on the difficulty level of the task: eight for the high difficulty and four for the low difficulty. 
In the competitive task, the purpose was to pass more checkpoints than the opponent, while in the cooperative task, the purpose was to pass checkpoints through cooperation. 
Other detailed conditions are explained in later sections.
\\ \indent
To examine only the factors of interest in this experiment, the appearance and behavior of the agents were also standardized. 
In addition, agents did not speak during the task but only made minimal gestures. 
Participants interacted with an agent in one of four conditions, combining task difficulty (high, low) and task content (competitive, cooperative).
\\ \indent
Afterwards, the participants' empathy toward the agent was aggregated by another questionnaire similar to before the task. 
Then, the participants were asked to write their impressions of the experiment in free form.

\subsubsection{Participants}
Participants were recruited for the experiment using the Yahoo! crowdsourcing company. 
Participants were paid 55 yen after completing all tasks as a reward for participating. 
A website was created for the experiment, which was limited to using a PC.
\\ \indent
There were a total of 596 participants.
However, there were 18 participants who gave inappropriate responses, which were eliminated as erroneous data, leaving a total of 578 participants. 
To judge whether answers were inappropriate in the experiment, we judged answers as inappropriate when the changes in the empathy values before and after the task were the same for all items or when only one item changed \cite{Schonlau15,Leiner19}. 
The task of aligning the number of participants to an appropriate number for the analysis was performed, and 142 participants in each condition were included in the analysis, starting from the top in the order of their participation in the experiment. 
Thus, the total number of participants used in the analysis was 568.
\\ \indent
The average age was 48.32 years (standard deviation 11.02), with a minimum of 18 years and a maximum of 87 years. 
The gender breakdown was 344 males and 224 females.

\subsubsection{Task difficulty}
Two task difficulty levels were prepared for this experiment. 
Figures 1 and 2 show these levels. 
The following conditions were used for different levels of difficulty.
\\ \indent
A) The total number of squares that could be moved to was eight for the high difficulty level and four for the low difficulty level. 
B) The number of checkpoints is seven, regardless of difficulty level.
C) In the case of the competitive task, the high difficulty level required the participant to act in such a way that at least four checkpoints were passed, while the low difficulty level required them to act in such a way that at least two checkpoints were passed. 
D) In the case of the cooperative task, for the high difficulty level, the human and agent had to cooperate to pass all seven checkpoints, and for the low difficulty level, they had to cooperate to pass at least four checkpoints.
\\ \indent
By reducing the total number of moves to be made by half and simplifying the expected number of checkpoints that the participants and agents had to pass through for the low-difficulty level, the number of trials was made to have a significant effect on the difficulty level.

\subsubsection{Task content} 
Two types of task content were prepared. 
By keeping the task environments as close as possible, we eliminated external factors and tried to measure the effect of task content on human empathy. 
The two types of task content were competitive and cooperative.
\\ \indent
In the competitive task, the task was a checkpoint competition, and the number of checkpoints required to win varied depending on the difficulty level. 
Points were awarded to the first person to pass each checkpoint. 
The win ratio for a total of three tasks was one win, one loss, and one tie, even when participants took the optimal actions. 
The win rate was adjusted to reduce the impact of the win rate on human empathy.
\\ \indent
In the cooperative task, the purpose of the task was for the participant and agent to pass all the checkpoints, and the total number of checkpoints varied with the difficulty level. 
The high difficulty level required the human and agent to cooperate to pass all seven checkpoints, and the low difficulty level required them to cooperate to pass at least four checkpoints. 
The maximum number of checkpoints that can be passed by each participant in a total of three tasks was adjusted. 
This was done to prevent one side from always passing too many odd-numbered checkpoints. 
It does not make sense if both parties pass through the same checkpoint.

\subsubsection{Questionnaire}
Participants completed a questionnaire before and after the task. 
The questionnaire was a 12-item questionnaire modified from the Interpersonal Reactivity Index (IRI), which is used to investigate the characteristics of empathy, to suit the present experiment \cite{Davis80}. 
The two questionnaires were the same. 
Both used were based on the IRI and were surveyed on a 5-point Likert scale (1: not applicable, 5: applicable). 
The questionnaire used is shown in Table 1. Q4, Q9, and Q10 are inverted items, so the scores were reversed when analyzing them.
\renewcommand{\arraystretch}{1.1}
\begin{table*}[tbp] 
    \caption{Summary of questionnaire used in study 1}
    \centering
    \scalebox{1.0}{
    \begin{tabular}{l}\hline 
        \textbf{Affective empathy}\\ \hline
        \textbf{Personal distress}\\
        Q1: If an emergency happens to the character, you would be anxious and restless.\\
        Q2: If the character is emotionally disturbed, you would not know what to do.\\
        Q3: If you see the character in need of immediate help, you would be confused and would not know what to do.\\
        \textbf{Empathic concern}\\
        Q4: If you see the character in trouble, you would not feel sorry for that character.\\
        Q5: If you see the character being taken advantage of by others, you would feel like you want to protect that character.\\
        Q6: The character's story and the events that have taken place move you strongly.\\\hline
        \textbf{Cognitive empathy}\\ \hline
        \textbf{Perspective taking}\\
        Q7: You look at both the character's position and the human position.\\
        Q8: If you were trying to get to know the character better, you would imagine how that character sees things.\\
        Q9: When you think you're right, you don't listen to what the character has to say.\\
        \textbf{Fantasy scale}\\
        Q10: You are objective without being drawn into the character's story or the events taken place.\\
        Q11: You imagine how you would feel if the events that happened to the character happened to you.\\
        Q12: You get deep into the feelings of the character.\\\hline
    \end{tabular}}
    \label{table1}
\end{table*}
\renewcommand{\arraystretch}{1.0}

\subsubsection{Analysis method}
The analysis was a two-factor analysis of variance (ANOVA). 
The between-participant factors were two levels of task difficulty and two levels of task content.
On the basis of the results of the participants' questionnaires, we investigated how task difficulty and task content influenced the promotion of empathy as factors that elicit human empathy. 
The numerical values of empathy aggregated before and after the task were used as the dependent variable.
R (R ver. 4.1.0) was used for the ANOVA. 
Also, we used anovakun (ver. 4.8.6) as the R package.

\section{Study 1 result}
All 12 questionnaire items were analyzed together. 
We also categorized and analyzed affective and cognitive empathy. 
For multiple comparisons, Holm's multiple comparison test was used to examine the existence of significant differences. 
Table 2 shows the results of the overall analysis. 
The results of the questionnaire analysis showed a main effect of task difficulty on affective empathy. 
The results are shown in Figure 4.
\renewcommand{\arraystretch}{1.1}
\begin{table*}[tbp]
    \caption{Results of all analyses of variance}
    \scalebox{1.0}{
        \begin{tabular}{c|c|c|cc|c|ccc}\hline 
        \multicolumn{2}{c|}{Category} & Conditions & Mean & S.D. & Factor & \em{F} & \em{p} & $\eta^2_p$\\ \hline 
        & & high-competitive & 38.7746 & 6.0675 & difficulty & 0.7877 & 0.3752 \em{ns} & 0.0014 \\ 
        & & high-cooperative & 38.4366 & 5.6949 & content & 0.6676 & 0.4142 \em{ns} &  0.0012 \\ 
        Empathy & pre & low-competitive & 38.4014 & 6.1009 & interaction & 0.0198 & 0.8880 \em{ns} & 0.0000 \\ 
        & & low-cooperative & 37.9225 & 5.9562 & & & & \\ \cline{2-9}
        & & high-competitive & 37.7535 & 6.5852 & difficulty & 2.4737 & 0.1163 \em{ns} & 0.0044 \\ 
        (Q1-Q12) & & high-cooperative & 37.8169 & 6.0980 & content & 0.1053 & 0.7456 \em{ns} & 0.0002 \\ 
        & post & low-competitive & 37.1127 & 7.3389 & interaction & 0.1909 & 0.6624 \em{ns} & 0.0003 \\ 
        & & low-cooperative & 36.6831 & 6.8098 & & & & \\ \hline
        & & high-competitive & 19.4859 & 3.6021 & difficulty & 2.2672 & 0.1327 \em{ns} & 0.0040  \\ 
        & & high-cooperative & 19.5352 & 3.6059 & content & 0.1374 & 0.7110 \em{ns} & 0.0002 \\ 
        Affective & pre & low-competitive & 19.1901 & 3.7282 & interaction & 0.2839 & 0.5944 \em{ns} & 0.0005 \\ 
        empathy & & low-cooperative & 18.9155 & 3.5520 & & & & \\ \cline{2-9}
        & & high-competitive & 18.8662 & 3.8798 & difficulty & 4.0986 & 0.0434 * & 0.0072 \\ 
        (Q1-Q6) & & high-cooperative & 18.9648 & 3.5479 & content & 0.1855 & 0.6668 \em{ns} & 0.0003\\ 
        & post & low-competitive & 18.4437 & 4.1642 & interaction & 0.5362 & 0.4643 \em{ns} & 0.0009 \\ 
        & & low-cooperative & 18.0634 & 3.9683 & & & & \\ \hline
        & & high-competitive & 19.2887 & 3.2586 & difficulty & 0.0027 & 0.9584 \em{ns} & 0.0000\\ 
        & & high-cooperative & 18.9014 & 3.0325 & content & 1.2025 & 0.2733 \em{ns} & 0.0021 \\ 
        Cognitive & pre & low-competitive & 19.2113 & 3.2109 & interaction & 0.1152 & 0.7344 \em{ns} & 0.0002 \\ 
        empathy & & low-cooperative & 19.0070 & 3.3464 & & & & \\ \cline{2-9}
        & & high-competitive & 18.8873 & 3.2900 & difficulty & 0.5918 & 0.4421 \em{ns} & 0.0010 \\ 
        (Q7-Q12) & & high-cooperative & 18.8521 & 3.4124 & content & 0.0208 & 0.8854 \em{ns} & 0.0000 \\ 
        & post & low-competitive & 18.6690 & 3.7088 & interaction & 0.0006 & 0.9808 \em{ns} & 0.0000 \\ 
        & & low-cooperative & 18.6197 & 3.5385 & & & & \\ \hline
        \multicolumn{9}{c}{}
        \end{tabular}} \\
            \em{p}:
{{*}p\textless\em{0.05}}
{{**}p\textless\em{0.01}}
{{***}p\textless\em{0.001}}
    \label{table2}
\end{table*}
\renewcommand{\arraystretch}{1.0}
\begin{figure}[tbp]
    \centering
    \includegraphics[scale=0.25]{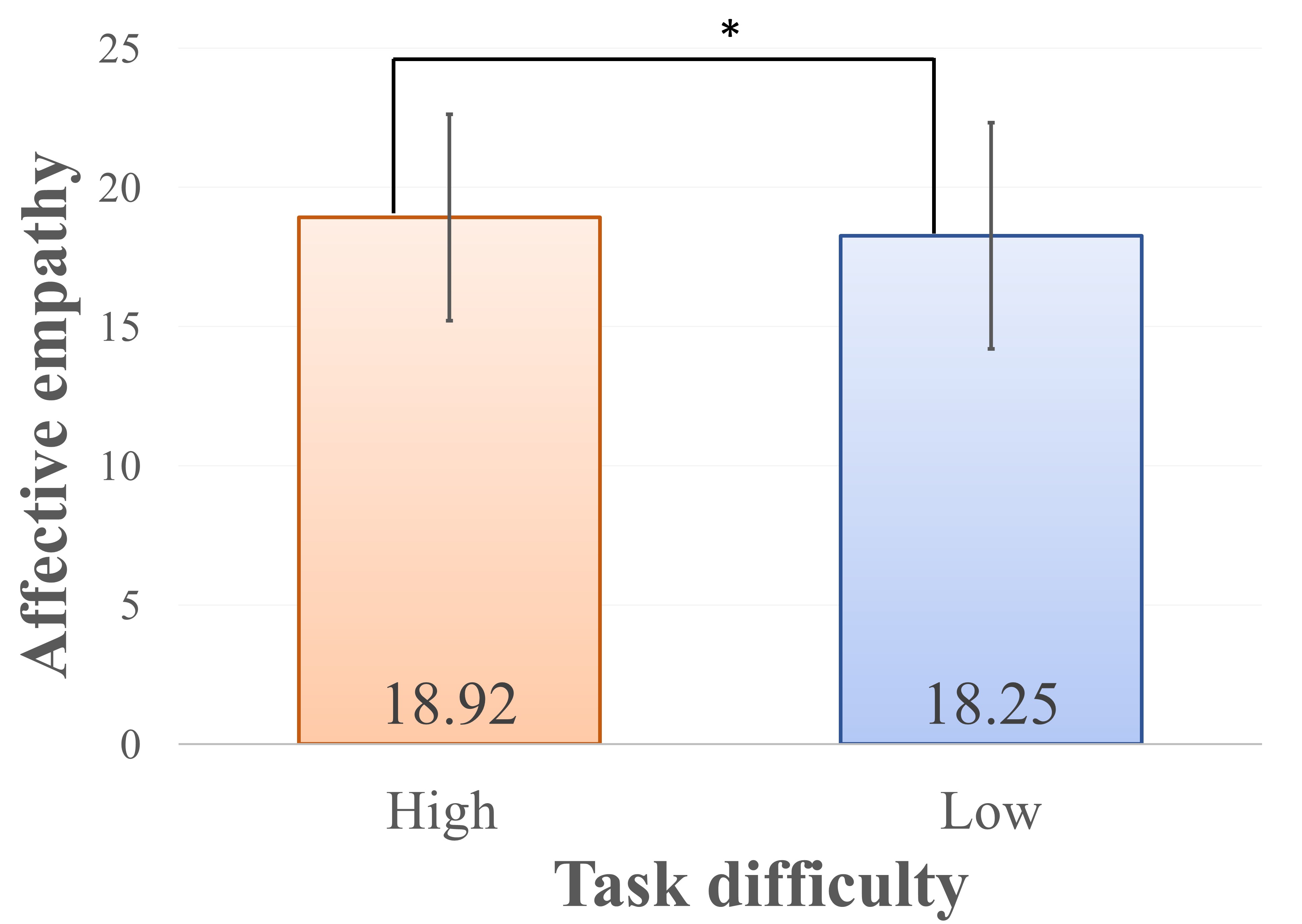}
    \caption{Main effects results of affective empathy}
    \label{fig4}
\end{figure}
\\ \indent
Initially, as can be seen from Table 2, we examined differences in empathy toward the agent among the participants and found no differences between any of the conditions. 
Therefore, we assumed that the ability to empathize with the agent was similar among the participants.
In this study, the questionnaire on pre-task empathy toward the agent was given only to confirm that there were no differences between participants. 
Therefore, we do not discuss significant differences between the pre- and post-task cases.
\\ \indent
The post-task results showed no interaction between task difficulty and task content, regardless of empathy category. 
For the 12 items, there was also no main effect of each factor [$F$(1,564) = 2.4737]. 
Similarly, no main effect of task content was found [$F$(1,564) = 0.5918]. 
However, a main effect was found for task difficulty [$F$(1,564) = 4.0986] based on the analysis of affective empathy in Table 2.
The main effect of post-task task-difficulty was higher for affective empathy for the higher difficulty level (high difficulty: mean = 18.9155, S.D. = 3.7113; low difficulty: mean = 18.2535, S.D. = 4.0647), as shown in Figure 4. 
On the basis of the above analysis, the results of this experiment suggest that a higher task difficulty promotes affective empathy.
\\ \indent
The post-task values were lower than the pre-task values of empathy toward the agent in each condition (all pre-task: mean = 38.3838, S.D. = 5.9490; all post-task: mean = 37.3415, S.D. = 6.7214). 
Also, for affective empathy, pre-task emotional empathy values were higher (all pre-task: mean = 19.2817, S.D. = 3.6216; all post-task: mean = 18.5845, S.D. = 3.9027). 
Similarly, for cognitive empathy, pre-task cognitive empathy values were higher (all pre-task: mean = 19.1021, S.D. = 3.2094; all post-task: mean = 18.7570, S.D. = 3.4835). 
\\ \indent

\section{Study 2 method}
In Study 1, human empathy for the agent did not strongly affect task difficulty and task content.
\color{black}
Therefore, in this study, we focus on an agent's expression as a way to increase the empathy felt by humans. 
Specifically, we experimentally investigate the possibility of an agent's expression facilitating human empathy. 
In addition, Study 1 did not aim to investigate changes in empathy before and after the task.
Study 2 defined the agent's expression as a factor other than appearance.
Therefore, in this experiment, pre- and post-task are considered as an additional factor.
\color{black}
Experiments were conducted utilizing a three-way mixed design in which the factors were the agents' expression (available, not available), task completion (success, failure), and empathy evaluation (before, after a task).

\subsubsection{Experimental Purpose and Design}
Our objective in this study is to determine whether the success or failure of a task and the agent's expression during the task elicits more human empathy as a result of the human's interactions with the agent. 
We designed a task based on a typing game and recruited participants for an experiment in which the empathy agent interacted during the task by expressing itself to the participants. 
We came up with the following two hypotheses.
\vspace{1mm}
\begin{enumerate}
\item[\textbf{H3}:] Successful completion of a task has less impact on human empathy than failure of a task.
\item[\textbf{H4}:] When an empathy agent makes an expression, it promotes empathy regardless of the success or failure of the task.
\end{enumerate}
\vspace{1mm}

We designed a between-participants experiment with two factors: task completion and agent expression. 
The number of levels for each factor was two for task completion (success, failure) and two for agent expression (available, not available). 
The agent's expression is a factor other than the agent's appearance. 
Two levels of pre- and post-task empathy values were used as within-participant factors. 
The dependent variable was the empathy that the participants felt.
\\ \indent
There are two reasons we focused on an agent's expression. 
The first is to investigate whether the agent's expression affects empathy even when the agent is not directly related to the task. 
The second is to investigate whether the agent's expression influences human motivation for the task. 
We also focused on task completion because the effect of the agent's expression could change depending on whether the task was a success or failure. 

\subsubsection{Experimental Details}
The experiment was conducted online using a PC. 
We utilized one of the experimental methods from an earlier study \cite{Davis99,Crump13,Okamura20}. 
Figure 5 shows the process flow.

\begin{figure*}[tbp]
\centering
\includegraphics[scale=0.35]{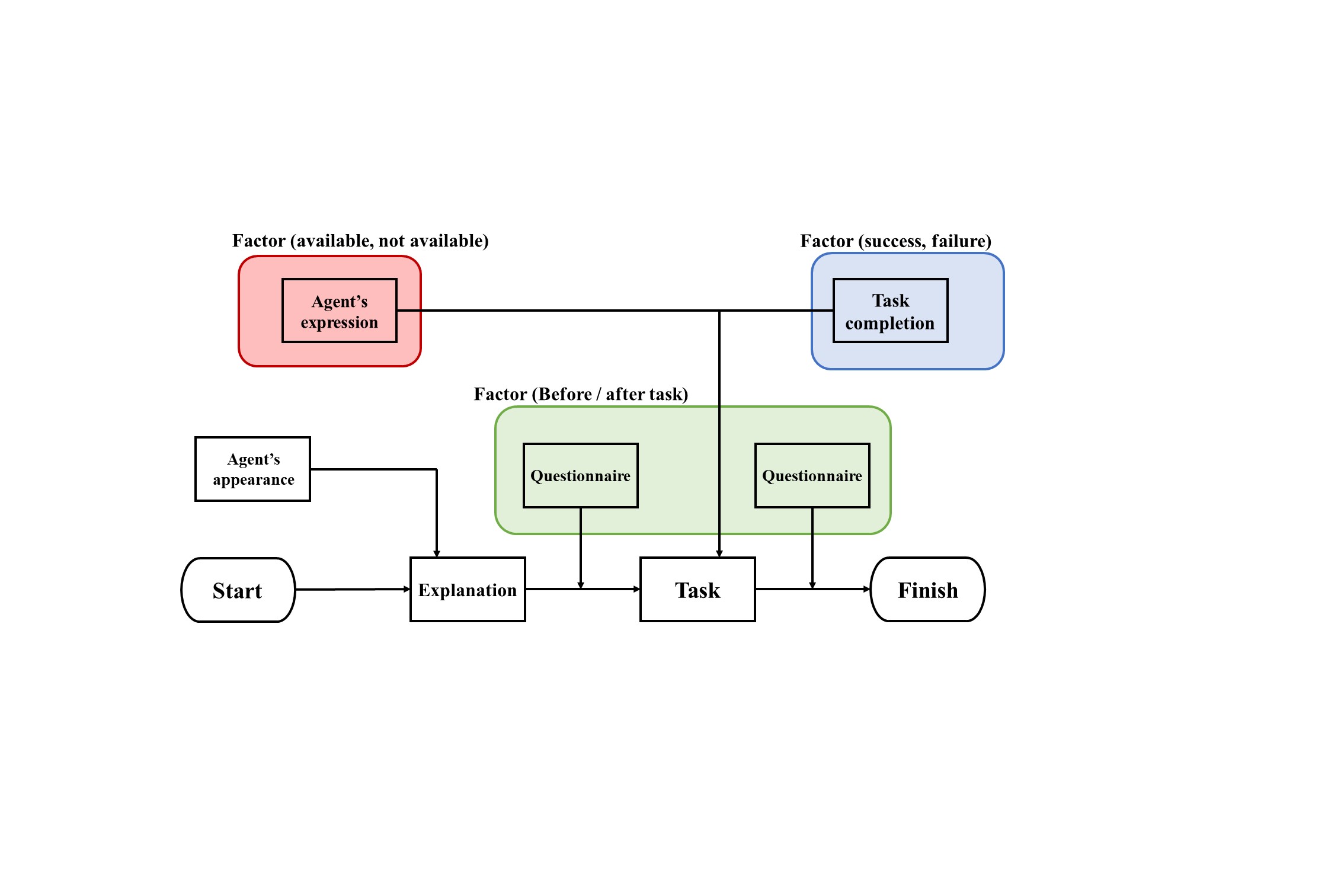}
\vspace{1mm}
\caption{Process flow of the experiment.}
\label{fig5}
\end{figure*}

As our primary goal is to promote human empathy toward anthropomorphic agents, we administered a questionnaire to measure empathy toward the agent before participants performed the task. 
This is the part of the explanation in Fig. 5. 
The participants were told that the agent had the role of watching over the task.
\\ \indent
The task was a typing game in which participants were asked to type 150 random letters of the alphabet. 
Our aim was to determine the standard typing time for the success or failure of the task, which is a factor in this experiment. 
\\ \indent
The agent expressed itself in various ways before and after the typing game started and ended. 
Specifically, before the start, it made cheering comments and gestures, and after the end, it made praising comments and gestures if the task was a success, and encouraging comments and gestures if the task was a failure. 
The agent did not make any gestures or expressions during the game so as not to affect the typing (i.e., it did nothing). 
After the task was completed, we administered the same questionnaire as before the task and additionally asked participants about their motivation to continue the task in order to investigate their empathic response. 
After completing the questionnaire, participants voluntarily answered the free-response questions.

\subsubsection{Participants}
Participants were recruited through Yahoo Crowdsourcing and paid 32 yen (= 0.22 dollars). 
The original number of participants was 398, but we excluded 35 who either gave inappropriate responses (i.e., the change in empathy values before and after the task was the same for all items, or only changed for one item \cite{Schonlau15,Leiner19}) or seemed unmotivated due to slow typing speed (less than one letter in two seconds). 
\\ \indent
We then applied Cronbach's $\alpha$ coefficient to the remaining 363 participants to determine the reliability of the questionnaire responses, which was found to be between 0.8027 and 0.9025 in all conditions. 
Sixty-eight participants in each condition were included in the analysis in the order of participation. 
At this time, the gender of the analyzed participants was adjusted so that there was no difference between conditions. 
Thus, the total number of participants used in the analysis was 272.
\\ \indent
The average age was 47.56 years (standard deviation 10.76), with a minimum of 19 years and a maximum of 75. 
The gender was 136 males and 136 females.

\subsubsection{Questionnaire}
Participants answered questionnaires before and after the task. 
This was a 12-item questionnaire modified from the Interpersonal Reactivity Index (IRI). 
As the IRI was designed to investigate human empathy characteristics, we modified it to investigate empathy toward an empathy agent. 
The same questionnaire was used both before and after the task and was administered on a 5-point Likert scale (1: not applicable, 5: applicable), as shown in Table 3. 
Q4, Q9, and Q10 are inverted items, so the scores were reversed when analyzing them.
\\ \indent
Q1 to Q6 examine affective empathy, and Q7 to Q12 examine cognitive empathy. 
There is one additional item for the questionnaire administered after the text (BeQ in the table) to examine the empathic response of the participants.

\renewcommand{\arraystretch}{1.1}
\begin{table*}[tbp] 
    \caption{Summary of questionnaire used in study 2}
    \vspace{1mm}
    \centering
    \scalebox{1.0}{
    \begin{tabular}{l}\hline 
        \textbf{Affective empathy}\\ \hline
        \textbf{Personal distress}\\
        Q1: If an emergency happens to the character, you would be anxious and restless.\\
        Q2: If the character is emotionally disturbed, you would not know what to do.\\
        Q3: If you see the character in need of immediate help, you would be confused and would not know what to do.\\
        \textbf{Empathic concern}\\
        Q4: If you see the character in trouble, you would not feel sorry for that character.\\
        Q5: If you see the character being taken advantage of by others, you would feel like you want to protect that character.\\
        Q6: The character's story and the events that have taken place move you strongly.\\\hline
        \textbf{Cognitive empathy}\\ \hline
        \textbf{Perspective taking}\\
        Q7: You look at both the character's position and the human position.\\
        Q8: If you were trying to get to know the character better, you would imagine how that character sees things.\\
        Q9: When you think you're right, you don't listen to what the character has to say.\\
        \textbf{Fantasy scale}\\
        Q10: You are objective without being drawn into the character's story or the events taken place.\\
        Q11: You imagine how you would feel if the events that happened to the character happened to you.\\
        Q12: You get deep into the feelings of the character.\\\hline
        \textbf{Empathic response}\\ \hline
        BeQ: Do you want to continue to do tasks with the character in the future?\\ \hline
    \end{tabular}}
    \label{table3}
\end{table*}
\renewcommand{\arraystretch}{1.0}

\subsubsection{Agent's expression}
We prepared three types of expression for the agent, as shown in Fig. 6. 
Note that we did not make any special considerations for the agent's appearance: it was simply made to look robotic, since we wanted to reduce any impression based on gender. 
\begin{figure*}[tbp]
\includegraphics[scale=0.3]{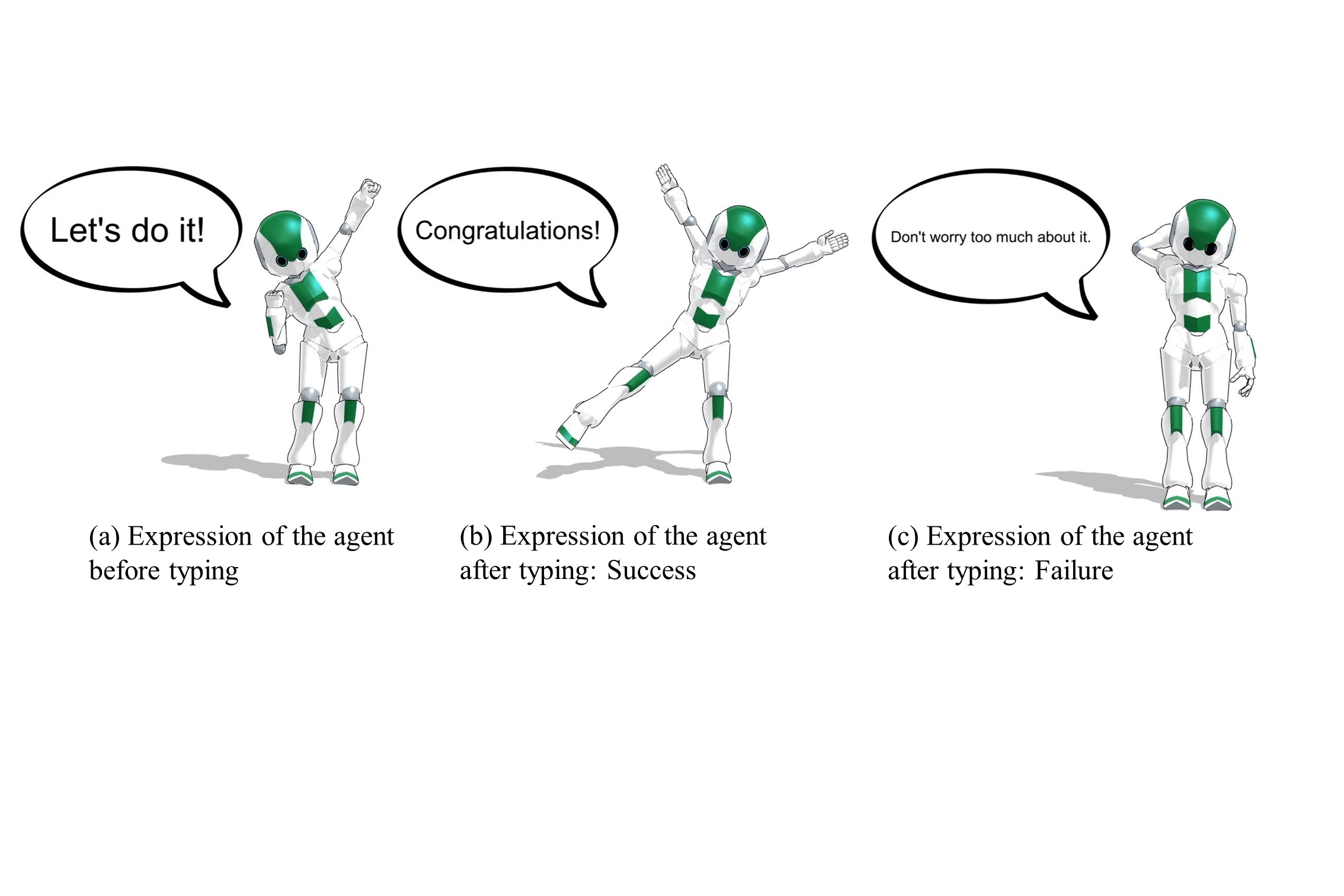}
\caption{Three expressions of the agent.}
\label{fig6}
\end{figure*}

Before the participant starts typing, the agent cheers in encouragement, as shown in Fig. 6(a). 
After typing, if the participant succeeds in the task, the agent expresses that it is pleased, as shown in 6(b).
If the participant fails, the agent makes a consoling expression, as shown in 6(c).
\\ \indent
The agent does not make any expressions while the participant is typing so as to allow him or her to concentrate on the task. 
It is standing in an upright state at this time. 
To confirm, participants are asked to indicate whether they noticed the agent's expression, as well as whether or not they were able to finish the task. 

\subsubsection{Task completion}
The experimental task is a typing game in which 150 random letters of the alphabet are entered. 
Since it is conducted online, there was a possibility that some participants might abandon the experiment if they knew there was a time limit. 
We therefore conducted the experiment without showing the time limit and simply terminated the task when the time was up. 
Figure 7 shows screenshots of the agent and the typing game during the experiment.

\begin{figure}[tbp]
\centering
\includegraphics[scale=0.4]{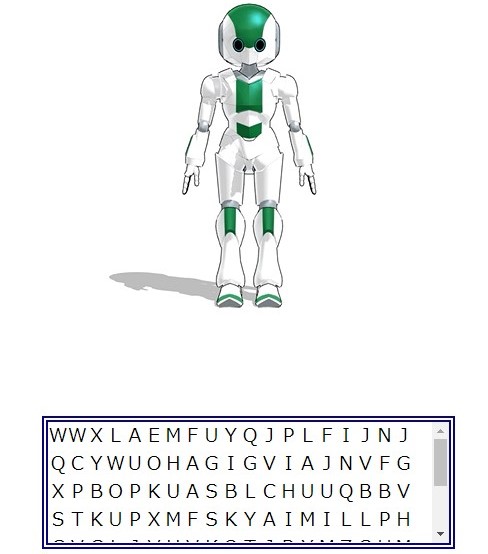}
\caption{Screenshots of the agent and the typing game during the experiment.}
\label{fig7}
\end{figure}

We conducted a pre-experiment to determine the amount of time to allow for successful completion of the task. 
A total of 50 participants were recruited for 20 yen (= 0.14 dollars). 
The mean age was 43.54 years (standard deviation 9.467) with a minimum of 22 years and a maximum of 65, and there were 46 men and four women. 
Results showed that the mean time to complete typing was 2 minutes and 4 seconds, with a standard deviation of 50 seconds, so the experiment was conducted with a standard deviation of $\pm{1}$ as typing success.

\subsubsection{Analysis Method}
The analysis was a three-factor mixed design analysis of variance (ANOVA). 
The between-participants factors were two levels of agent expression (available, not available) and two levels of task completion (success, failure). 
The within-participants factor consisted of two levels of empathy values before and after the task.
\\ \indent
With the results of the participants' questionnaire responses as a basis, we investigated the influence of agent expression and task completion as factors that elicit human empathy. 
The values of empathy before and after the task were used as the dependent variable. R (ver. 4.1.0) was used for the ANOVA.

\section{Study 2 result}
We analyzed the questionnaire responses using ANOVA and classified empathy into affective and cognitive empathy categories. 
For multiple comparisons, Holm's multiple comparison test was used to examine the existence of significant differences. 
The ANOVA results are shown in Table 4, where we can see there was an interaction between the two factors of agent expression and the pre- and post-task. 
The results of our analysis of the interaction are shown in Fig. 8, which also shows the mean and standard deviation for each condition.

\renewcommand{\arraystretch}{1.1}
\begin{table*}[tbp]
\caption{Analysis results of ANOVA}
\scalebox{1.0}{
\begin{tabular}{cllll}\hline
& \multicolumn{1}{c}{Factor} & \multicolumn{1}{c}{\em{F}} & \multicolumn{1}{c}{\em{p}} & \multicolumn{1}{c}{$\eta^2_p$}\\ \hline
& Agent's expression & 2.111 & 0.1474 \em{ns} & 0.0078 \\ 
& Task completion & 2.695 & 0.1019 \em{ns} & 0.0100\\
Empathy & Before/after task & 22.99 & 0.0000 *** & 0.0790 \\ 
(Q1-12)& Agent's expression $\times$ Task completion & 1.820 & 0.1785 \em{ns} & 0.0067 \\ 
& Agent's expression $\times$ Before/after task & 15.28 & 0.0001 *** & 0.0539 \\ 
& Task completion $\times$ Before/after task & 0.4515 & 0.5022 \em{ns} & 0.0017 \\
& Agent's expression $\times$ Task completion $\times$ Before/after task & 0.0233 & 0.8787 \em{ns} & 0.0001 \\ 
\hline
& Agent's expression & 2.164 & 0.1424 \em{ns} & 0.0080 \\ 
& Task completion & 2.941 & 0.0875 \em{ns} & 0.0109\\
Affective & Before/after task & 20.96 & 0.0000 *** & 0.0725 \\ 
empathy & Agent's expression $\times$ Task completion & 0.9994 & 0.3184 \em{ns} & 0.0037 \\ 
(Q1-6)& Agent's expression $\times$ Before/after task & 5.400 & 0.0209 * & 0.0198 \\ 
& Task completion $\times$ Before/after task & 0.4452 & 0.5052 \em{ns} & 0.0017 \\
& Agent's expression $\times$ Task completion $\times$ Before/after task & 0.2335 & 0.6294 \em{ns} & 0.0009 \\ 
\hline
& Agent's expression & 1.429 & 0.2330 \em{ns} & 0.0053 \\ 
& Task completion & 1.663 & 0.1984 \em{ns} & 0.0062\\
Cognitive & Before/after task & 10.90 & 0.0011 ** & 0.0391 \\ 
empathy & Agent's expression $\times$ Task completion & 2.353 & 0.1262 \em{ns} & 0.0087 \\ 
(Q7-12)& Agent's expression $\times$ Before/after task & 19.80 & 0.0000 *** & 0.0688 \\ 
& Task completion $\times$ Before/after task & 0.1854 & 0.6671 \em{ns} & 0.0007 \\
& Agent's expression $\times$ Task completion $\times$ Before/after task & 0.7919 & 0.3743 \em{ns} & 0.0029 \\ 
\hline
Empathic & Agent's expression & 12.52 & 0.0005 *** & 0.0446 \\ 
response & Task completion & 1.738 & 0.1885 \em{ns} & 0.0064\\ 
(BeQ) & Agent's expression $\times$ Task completion & 0.8136 & 0.3679 \em{ns} & 0.0030 \\ 
\hline
\multicolumn{5}{c}{}
\end{tabular}} \\
            \em{p}:
{{*}p\textless\em{0.05}}
{{**}p\textless\em{0.01}}
{{***}p\textless\em{0.001}}
\label{table4}
\end{table*}
\renewcommand{\arraystretch}{1.0}

\begin{figure*}[tbp]
    \includegraphics[scale=0.25]{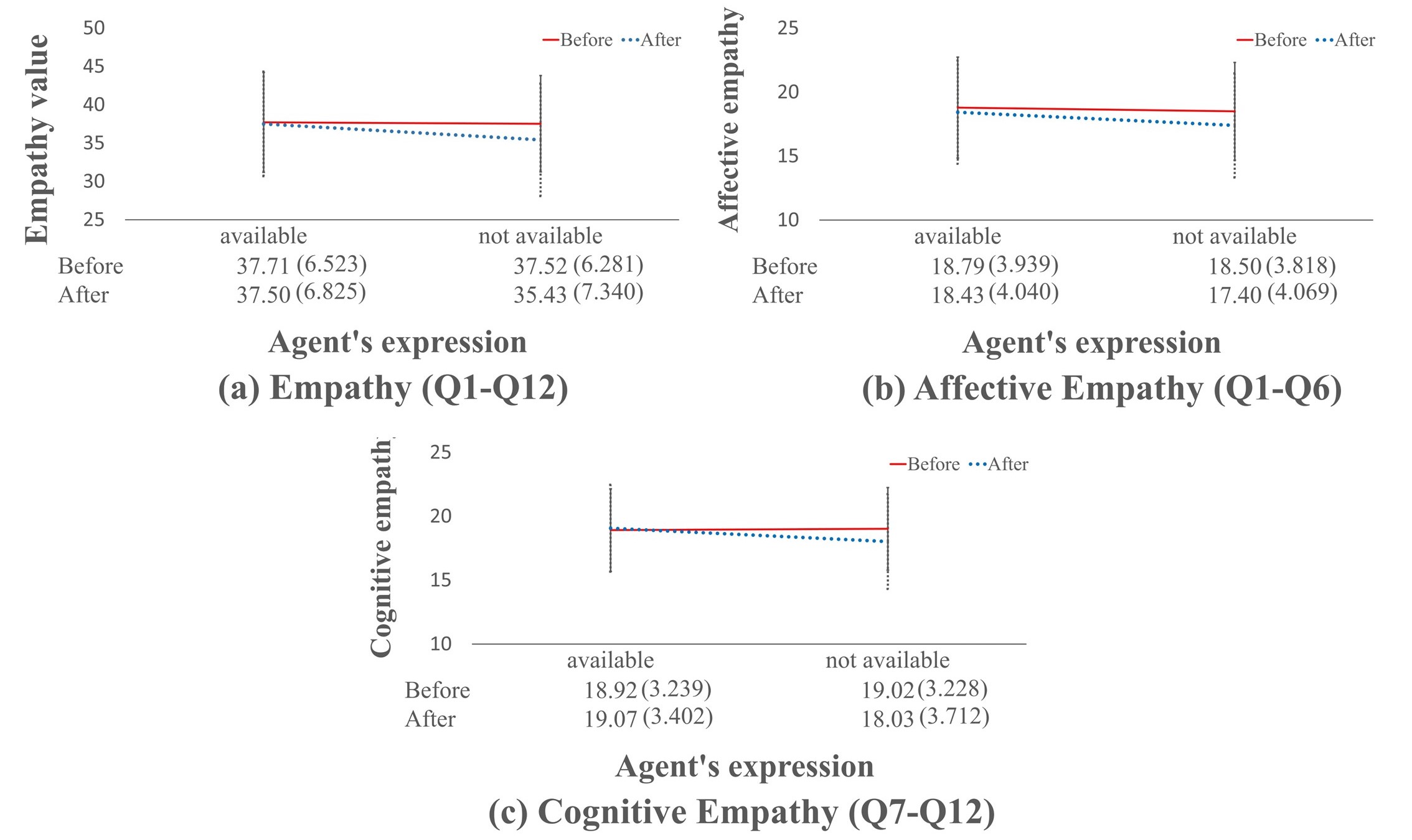}
    \caption{Interaction results for (a) empathy, (b) affective empathy, and (c) cognitive empathy.}
    \label{fig8}
\end{figure*}

No significant interactions between the agent's expression and task completion were found in all conditions. 
In the following, we omit a main effect's explanation of items for which a main effect and interaction were found. 
For items for which no interaction was found and a main effect was found, the results for the main effect are presented. 
Therefore, since an interaction between an agent expression and the pre- and post-task was found, the results of the analysis of the simple main effect are shown in Table 5.

\renewcommand{\arraystretch}{1.1}
\begin{table*}[tbp]
\caption{Analysis results of multiple comparison (simple main effect)}
\scalebox{1.0}{
\begin{tabular}{cllll}\hline
& \multicolumn{1}{c}{Factor} & \multicolumn{1}{c}{\em{F}} & \multicolumn{1}{c}{\em{p}} & \multicolumn{1}{c}{$\eta^2_p$}\\ \hline
& Agent's expression at before task & 0.0613 & 0.8046 \em{ns} & 0.0002 \\ 
Empathy & Agent's expression at after task & 5.849 & 0.0163 * & 0.0214\\ 
(Q1-12) & Before/after task at agent's expression available & 0.5485 & 0.4602 \em{ns} & 0.0041 \\
& Before/after task at agent's expression not available & 29.48 & 0.0000 *** & 0.1803 \\ 
\hline
Affective & Agent's expression at before task & 0.3951 & 0.5302 \em{ns} & 0.0015 \\ 
empathy & Agent's expression at after task & 4.454 & 0.0357 * & 0.0163\\ 
(Q1-6) & Before/after task at agent's expression available & 3.861 & 0.0515 \em{ns} & 0.0280 \\
& Before/after task at agent's expression not available & 17.76 & 0.0000 *** & 0.1170 \\ 
\hline
Cognitive & Agent's expression at before task & 0.0693 & 0.7925 \em{ns} & 0.0003 \\ 
empathy & Agent's expression at after task & 5.806 & 0.0166 * & 0.0212\\ 
(Q7-12) & Before/after task at agent's expression available & 0.7005 & 0.4041 \em{ns} & 0.0052 \\
& Before/after task at agent's expression not available & 28.36 & 0.0000 *** & 0.1747 \\ 
\hline
\multicolumn{5}{c}{}
\end{tabular}} \\
            \em{p}:
{{*}p\textless\em{0.05}}
{{**}p\textless\em{0.01}}
{{***}p\textless\em{0.001}}
\label{table5}
\end{table*}
\renewcommand{\arraystretch}{1.0}

\subsubsection{Empathy}
The results for empathy (Q1-Q12) revealed an interaction between the agent's expression and the pre- and post-task factors. 
The main effects of pre- and post-task were also significant, but were omitted because of the interaction between the agent's expression and the pre- and post-task factors.
\\ \indent
Multiple comparisons revealed a significant difference in the simple main effect of the post-task agent's expression factor, as shown in Fig. 9. 
In addition, significant differences were found in the simple main effects of the pre- and post-task factors in the non-expression condition of the agent. 
These results suggest that empathy was induced in the task with the agent's expression, while it was suppressed in the task without the agent's expression. 
The results of post-hoc analysis indicate that agent expression is effective in inducing empathy.

\begin{figure}[tbp]
\includegraphics[scale=0.23]{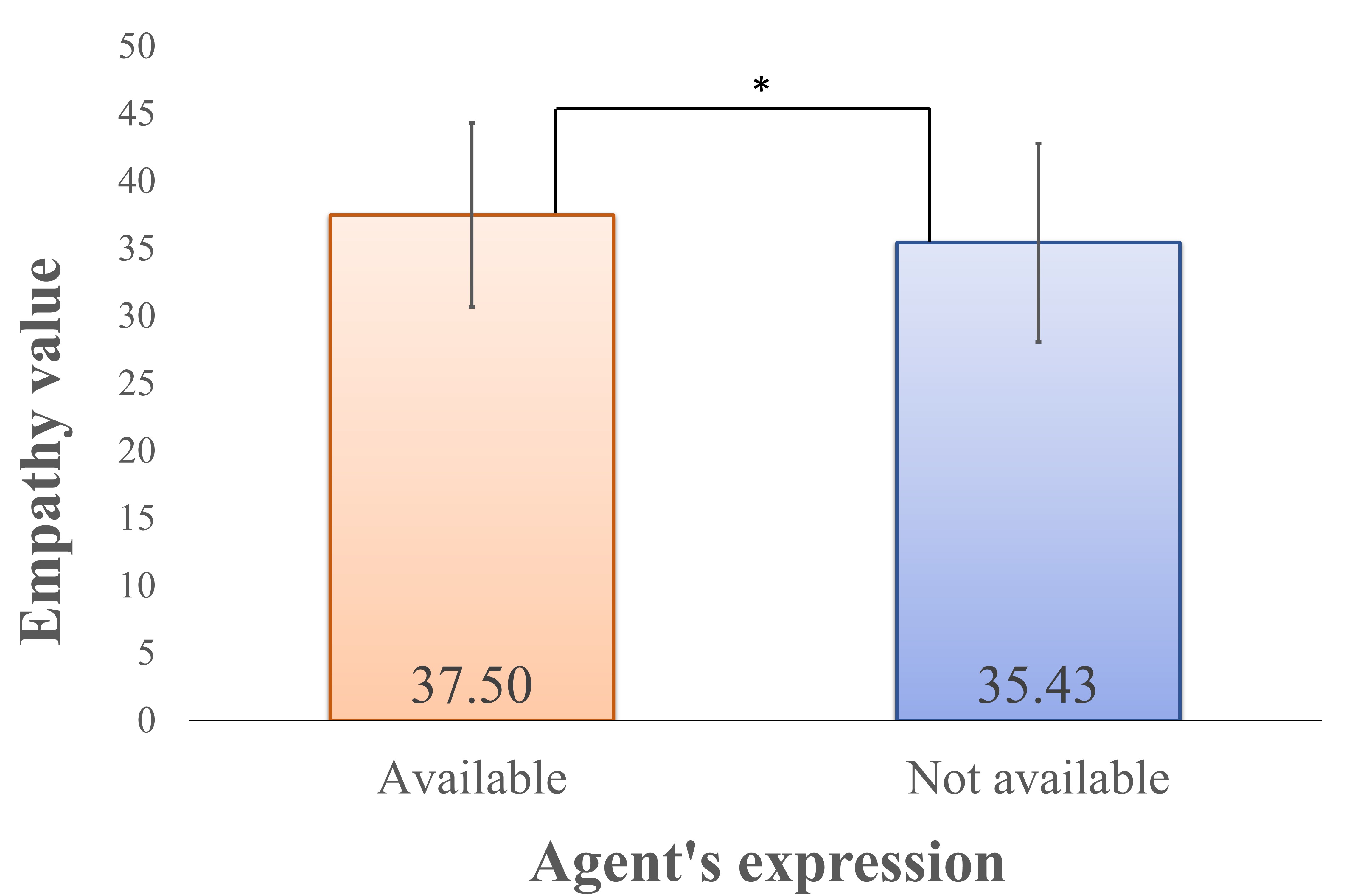}
\caption{Results of multiple comparisons of post-task empathy. Error bars show standard deviation.}
\label{fig9}
\end{figure}

\subsubsection{Affective empathy}
The results for affective empathy (Q1-Q6) revealed an interaction between the agent's expression and the pre- and post-task factors, the same as with empathy. 
The main effect of pre- and post-task was also significant, but was omitted because of the interaction between the agent's expression and the pre- and post-task factors.
\\ \indent
Multiple comparisons showed a significant difference in the simple main effect of the post-task agent's expression factor, as shown in Fig. 10. 
In addition, significant differences were found in the simple main effects of the pre- and post-task factors in the non-expression condition of the agent. 
These results suggest that affective empathy is induced in the presence of the agent's expression, while it is suppressed in the presence of the agent's non-expression. 
The results of post-hoc analysis indicate that the agent's expression is effective in inducing affective empathy.

\begin{figure}[tbp]
\includegraphics[scale=0.23]{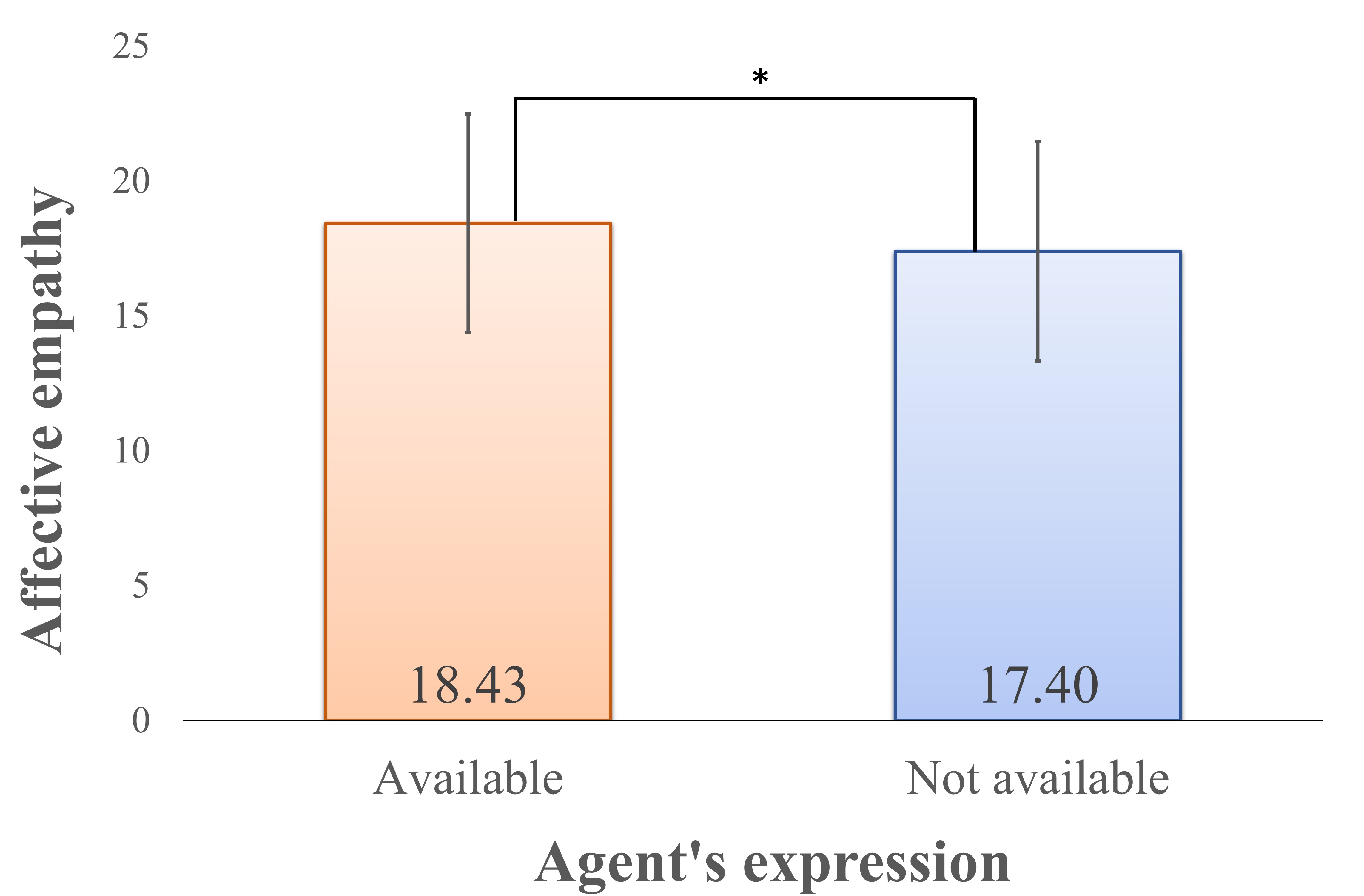}
\caption{Results of multiple comparisons of post-task affective empathy. Error bars show standard deviation.}
\label{fig10}
\end{figure}

\subsubsection{Cognitive empathy}
The results for cognitive empathy (Q7-Q12) showed an interaction between agent expression and the pre- and post-task factors, as was the case for empathy. 
The main effect of pre- and post-task was also significant, but was omitted because of the interaction between the agent's expression and the pre- and post-task factors.
\\ \indent
Multiple comparisons revealed a significant difference in the simple main effect of the post-task agent's expression factor, as shown in Fig. 11. 
In addition, significant differences were found in the simple main effects of the pre- and post-task factors in the non-expression condition of the agent. 
These results suggest that cognitive empathy is induced in the presence of the agent's expression, while it is suppressed in the presence of the agent's non-representation. 
The results of post-hoc analysis indicate that the agent's expression is effective in inducing cognitive empathy.

\begin{figure}[tbp]
\includegraphics[scale=0.23]{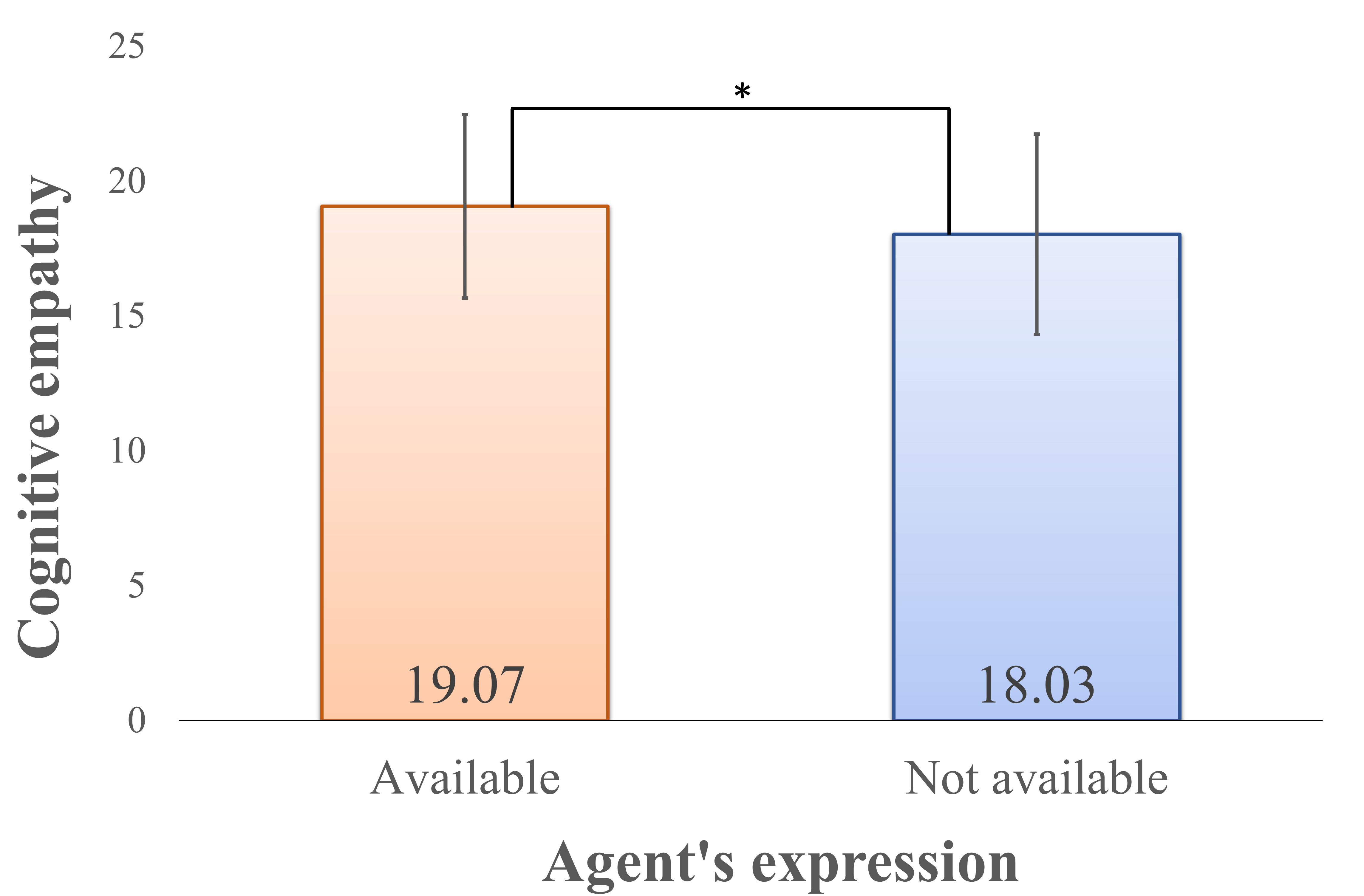}
\caption{Results of multiple comparisons of post-task cognitive empathy. Error bars show standard deviation.}
\label{fig11}
\end{figure}

\subsubsection{Empathic response}
There was no interaction between the agent's expression and task completion in the empathic response results. 
There was also no main effect of task completion.
\\ \indent
However, the main effect of the agent's expression was significant at two levels, as shown in Fig. 12, which indicates that the agent's expression can increase the participants' motivation for the task.

\begin{figure}[tbp]
\includegraphics[scale=0.23]{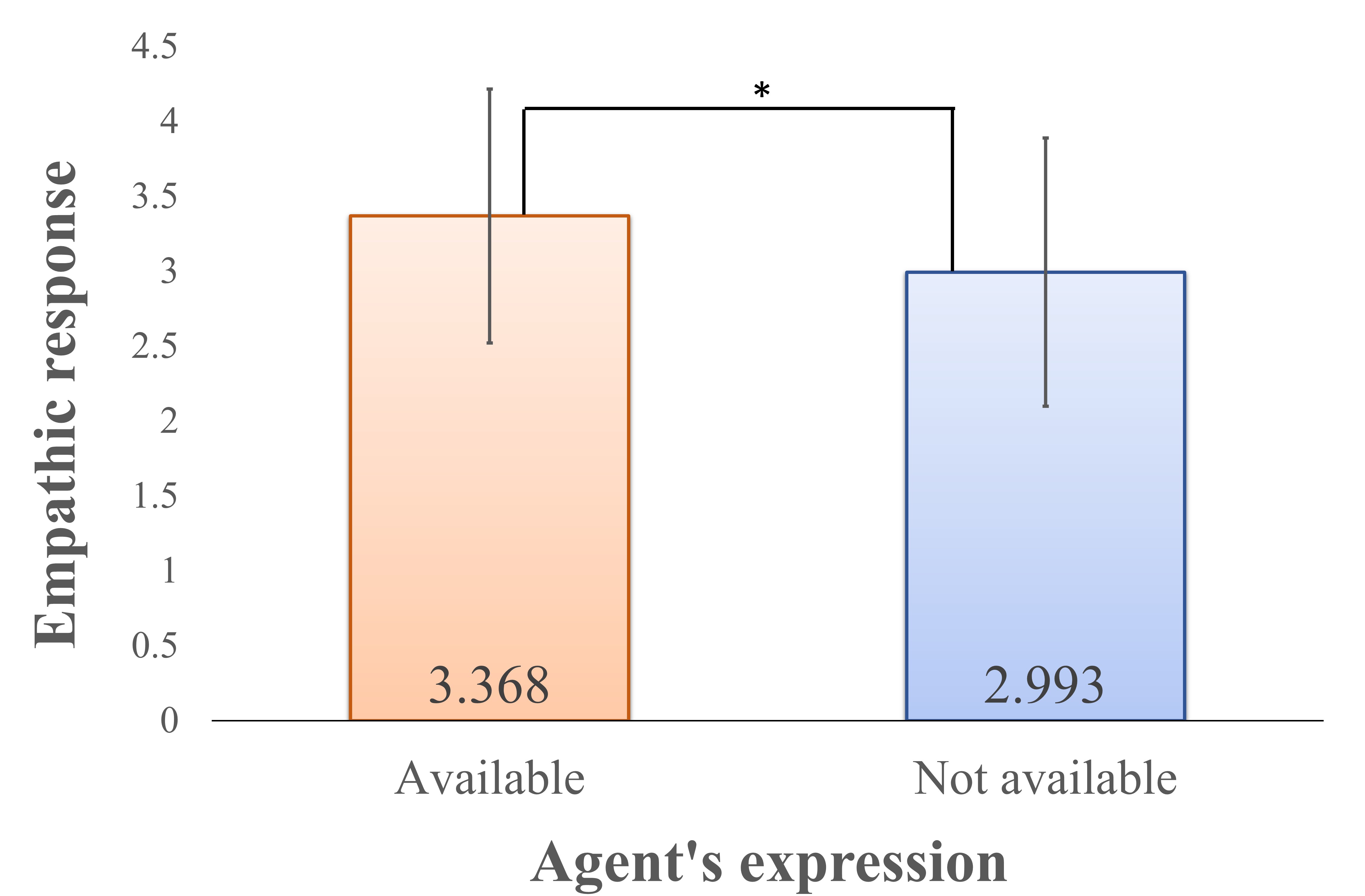}
\caption{Results of empathic response. Error bars show standard deviation.}
\label{fig12}
\end{figure}

\section{Discussion}
\subsection{Supporting Hypotheses}
Our concept for improving the relationship between humans and anthropomorphic agents is to get humans to empathize with the agents, which is an approach supported by several previous studies \cite{Gaesser13,Klimecki16}. Human empathy for agents is essential in order for agents to be utilized in society. If agents can take an appropriate approach to eliciting human empathy, humans and agents should be able to have a better relationship.

\subsubsection{Study 1}
This experiment was conducted to investigate the conditions necessary for a human to develop empathy for an anthropomorphic agent. 
In particular, this experiment aimed to identify factors that influence the empathy between an agent and a person who performed a task by investigating factors related to the task. 
For this purpose, we formulated the following two hypotheses and analyzed the data obtained from the experiment. 
\\ \indent
The results did not support H1 and H2. 
However, the results showed that regardless of task content, a higher task difficulty promoted human affective empathy. 
In discussing these results, it is necessary to focus on the changes in empathy before and after the task. 
\\ \indent
If only post-task surveys had been conducted, a higher task difficulty would have been shown to increase human empathy. 
However, by conducting a pre-task survey, it was found that empathy actually decreased throughout the task. 
This result indicated that post-task changes do not necessarily lead to better results than pre-task changes.

\subsubsection{Study 2}
We conducted our experiment to clarify the conditions necessary for humans to empathize with anthropomorphic agents. Specifically, we investigated the possibility of influencing the empathy that humans have for agents by means of the agent's expression and the success or failure of a task as humans and agents interact during a task. We proposed two hypotheses and analyzed the data obtained from the experiment to see if they were supported.
\\ \indent
Our findings showed that H3, “Successful completion of a task has less impact on human empathy than failure of a task,” was not supported, as the experimental results did not show that task completion had any effect on human empathy. Success and failure are important factors in the task, and we predicted that participants' empathy for the agent might differ depending on the outcome, but no significant differences were found. One possible reason for this could be that participants performed the experimental task without knowing what the time limit was. This might have caused them not to empathize due to the sudden end of the task.
\\ \indent
In contrast, H4, “When an empathy agent makes an expression, it promotes empathy regardless of the success or failure of the task,” was supported. Although the agent did not do anything special during the typing process, its expression before and afterwards resulted in a maintained human empathy regardless of the task outcome. This result indicates that the agent's expression was effective in inducing and maintaining human empathy regardless of task outcome. 

\subsection{The influence of the task and the importance of the agent's expression}
In Study 1, we hypothesized from Ruissen and de Bruijn \cite{Ruissen16} and Fuentes-García et al. \cite{Fuentes19}. 
The results were similar to related studies in that a high task difficulty increased human empathy, although H1 and H2 were not supported. 
However, a comparison of pre- and post-task human empathy toward the agent showed that post-task empathy decreased. Therefore, we found that the task may have a negative effect on human empathy toward an agent.
\\ \indent
In Study 2, in addition to investigating differences in task completion, we also focused on pre- and post-task differences to see the influence of agent expression, which was not the case in Study 1. 
The results showed that, as in Study 1, human empathy toward the agent decreased statistically significantly in the absence of agent expression. 
On the other hand, human empathy toward the agent was maintained in the presence of agent expression. 
Studies by Beck et al. \cite{Beck10} and Deshmukh et al. \cite{Deshmukh18} have shown that agent expressions (gestures and comments) can enhance understanding from humans, and our study also elicited human empathy toward the agent. 
\\ \indent
Our study showed that when a human and an agent perform a task, the minimum necessary behaviors and words are not enough to elicit empathy from the human. 
However, adding subtle expressions from the agent can help maintain empathy from humans.

\subsection{Empathic response}
In study 2, participants played the role of observer of the empathic agent. 
We found that the observers (participants) showed empathic responses to information from the target (appearance, comments, gestures) and to information related to the task (success/failure).
\\ \indent
One of our objectives was to determine whether the participants would want to perform the task with the agent in the future, which we measured by analyzing their empathic response behavior. 
Our analysis revealed a main effect in the agent's expression indicating that when humans perform a simple task, they can maintain their motivation for the task if there is an agent by their side. 
This finding should prove beneficial for improving motivation in simple tasks.

\subsection{Limitations}
\subsubsection{Study 1}
As a limitation of this experiment, it is possible that the simplicity of the task itself may have decreased empathy because the task itself was perceived as tedious, due to eliminate factors other than task difficulty and task content.
\\ \indent
However, the fact that task difficulty did not affect the task content but affected human affective empathy may be an effective factor in controlling human empathy when empathic agents coexist in human society in the future. 
By setting the task difficulty appropriately, it is possible to maintain an appropriate distance from an agent without making the participant empathize more than necessary.
\\ \indent
The results of the analysis, which classified empathy into affective empathy and cognitive empathy, showed a main effect of task difficulty on affective empathy. 
The main reason for this main effect is thought to be that affective empathy increased as task difficulty increased due to the increased mental load caused by the task. 
No main effects of task difficulty or task content were observed for cognitive empathy. 
This is because cognitive empathy requires imagining the thoughts and feelings of others in terms of oneself and imagining them from the other's point of view, so the task in this experiment did not enhance cognitive empathy.

\subsubsection{Study 2}
The main limitation of this study is that participants did not know the time limit. 
We specified this condition in order to avoid cheating in the online experimental environment. 
When people are actually performing tasks in the real world, the completion deadline is almost always made known. 
Our future work will therefore investigate whether knowing the time limit affects participants’ task completion.
\\ \indent
In addition, we used textual comments and gestures as the agent's expression in this study, and did not include voice. 
It is not possible to control voice in an online environment, which is why we kept it silent. 
Future work should investigate which has more of an influence on empathy in the agent's expression: voice or text.
\\ \indent
Finally, we investigated participants' motivation for the task as an empathic response, and while we found that motivation was increased when an agent's expression occurred, it is impossible to know whether a continuous relationship was actually established without a long-term experiment. 
Therefore, it will be necessary to conduct additional experiments in which duration is included as a factor.

\section{Conclusion}
If anthropomorphic agents are to become more acceptable to humans, thus enabling their integration into society, it is essential that humans need to be able to empathize with the agents. 
This study is part of a large body of research focused on the factors that influence empathy between humans and agents. 
We conducted two studies to investigate whether the task affects human empathy toward an agent.
\\ \indent
Study 1 focused on human-agent tasks as part of the factors and conditions that make humans empathize with anthropomorphic agents, and it investigated task difficulty and task content. 
Study 2 investigated the agent's expression and task completion, on the basis of the results from Study 1. 
The dependent variable in both studies was the participants' empathy value toward the agent.
\\ \indent
The analysis in Study 1 revealed that a higher task difficulty increased emotional empathy after the task. 
Our findings in Study 2 showed that there was no main effect for the task completion factor and that when the agent's expression was present, more human empathy was induced. 
In addition, the agent's expression was found to affect the participants' motivation for the task. 
\\ \indent
This study provides a beneficial example of the effectiveness of changes in human empathy in terms of human-agent relationships. 
In the case of simple tasks, human empathy decreases in the absence of the agent's expression, while the presence of the agent's expression maintains human empathy. 
Thus, the agent's expression affected the motivation for the task. 
Future research could explore empathy agents in a variety of situations by strengthening or weakening specific empathy components.

\bibliographystyle{IEEEtran}
\bibliography{sample}

\EOD

\end{document}